# Visibility of capillaries in turbid tissues: an analytical approach

Gennadi Saiko[1] and Alexandre Douplik[2]
*Department of Physics, Toronto Metropolitan University, Toronto, Canada*
*gsaiko@ryerson.ca*

**Abstract**

Purpose: Visualization and monitoring of capillary loops in dermis and mucosa are interesting for various clinical applications, including rheumatology, early cancer, and shock detection. However, the limitations of existing imaging technologies are not well understood. Therefore, this study aimed to elucidate peculiarities of the subsurface defect visualization in realistic skin imaging geometries.

Methods: We used a perturbation approach for the light propagation in turbid tissues with mismatched boundaries. Defects were considered as negative light sources immersed in homogeneous media, which was described using diffuse approximation. The contrast ratio was used as an image quality metric.

Results: We have developed the single point subsurface defect model and extended it to horizontally- and vertically-arranged linear inhomogeneities. In particular, we have obtained explicit analytical expressions for the single point defect and the infinite linear defect buried at a certain depth (horizontally-arranged), which allows direct experimental verification.

Conclusions: The developed approach can be used for quick rough estimates while designing and optimizing imaging systems.

**Declarations**

**Funding** :

Not Applicable

**Conflicts of interest/Competing interests**

The authors declare no conflict of interest

**Availability of data and material**

Not Applicable

**Code availability**

Not Applicable

---

[1] https://orcid.org/ 0000-0002-5697-7609
[2] https://orcid.org/ 0000-0001-9948-9472

1. **Introduction**

Visualization and monitoring of capillary loops in dermis and mucosa are interesting for various clinical applications, including rheumatology, early cancer, and shock detection [1]. For example, unusual capillary loop shapes and capillary loop density can be precursors of cancer transformations (e.g., angiogenesis) or auto-immune diseases (scleroderma). In addition, rapid changes in their shape and sizes can be one of the first signs of shock development (circulatory collapse). These clinical applications drive continuous improvements in imaging techniques, including traditional lens-based optical systems and novel lensless [2] optical systems.

While certain techniques (like optical coherence tomography, OCT) can be used to visualize and quantify subsurface objects, there is a need for simpler, wide-field imaging techniques for high throughput screening of large skin/mucosa areas. With the advances in CMOS cameras, the tissue can be visualized with high resolution using inexpensive smartphone cameras, providing a significant value in telehealth settings. However, due to the low-contrast nature of images of absorption inhomogeneities (e.g., capillaries) in highly-light-scattering biotissues, there are substantial limitations on the widespread clinical use of such technologies.

Several techniques were proposed to increase this contrast. For example, it is known [3] that narrow-band imaging provides better contrast than white light imaging. Imaging contrast can also be increased by optical clearing (by decreasing scattering) or transforming the image into a different colorspace (e.g., RGB->HSV) [4]. For example, Goffredo et al. [5] considered various color channel transformations to increase sensitivity and specificity for such defects discovery.

In parallel with these continuous efforts to improve imaging techniques, there are attempts to understand their limitations better. In particular, one area of interest is to understand the maximum detectability depth for various subsurface inhomogeneities. This topic has been studied analytically [6], numerically [7], and experimentally [8].

A mathematical model for visualization of a point defect in a slab of the turbid tissue with mismatched boundary conditions was developed in [9]. However, the validity of usage of the point defect to mimic the blood vessels in the skin requires further analysis. Realistically, capillaries are linear objects arranged either vertically (in most parts of the skin) or horizontally (in some parts of the skin, e.g., nail folds). So, it could be helpful to extend the approach developed in [9] to other defects geometries, particularly linear ones. This study aimed to elucidate peculiarities of the subsurface defect visualization in realistic skin imaging geometries. We will develop an analytical approach to calculate the contrast ratio of individual point defects and then extend it to linear subsurface defects arranged vertically (normal capillaries) and horizontally (nailfold capillaries).

2. **Methods**

*2.1 Tissue model*

Human skin and mucosal tissues have a layered structure [10]. Based on our primary task to visualize the capillary grid, we can group covering tissues into (I) bloodless epithelium, (II) blood-containing papillary layer of the dermis (skin), or lamina propria (mucosa), and (III) underlying tissues (see Fig 1A). Living cells in epithelium receive oxygen and nutrients through the diffusion from capillaries located in the papillary layer underneath. Thus, the thickness of living cells epithelium layers is limited by the diffusion length of the oxygen and typically does not exceed 100µm. However, the stratum corneum, which includes "non-supplied" cells, can be thicker in some organs, such as feet soles or palms.

*2.2 Geometry*

Based on our tissue model, the epithelium (including stratum corneum) can be considered an optical filter that covers absorption features and deteriorates the quality of the image. To evaluate how the measured contrast ratio is affected by the presence of this outermost layer, we can consider the following model (see Fig 1B): the homogeneous top layer (Layer I) covers Layer II, which consists of 2 areas: a) homogeneous background, b) capillaries, which can be considered as heterogeneous (either absorption or scattering) features or "defects." Below this layer II, there is another layer III, which represents all underlying tissues. As we are interested in estimating the effects of the outermost surface layer, in simplifying calculations, we can consider simplified geometry (Fig 1C): the homogeneous semi-infinite tissue characterized by an absorption coefficient $\mu_a$ and "defects" described by the volume $V$ and absorption coefficient $\mu_a + \delta\mu_a$ and located at the depth $Z$ (here $\delta\mu_a$ is an incremental absorption coefficient associated with the defect). In this case, the contrast presented by the features will be maximized, and we can estimate the upper bound (the best-case scenario) for visualization of these particular features.

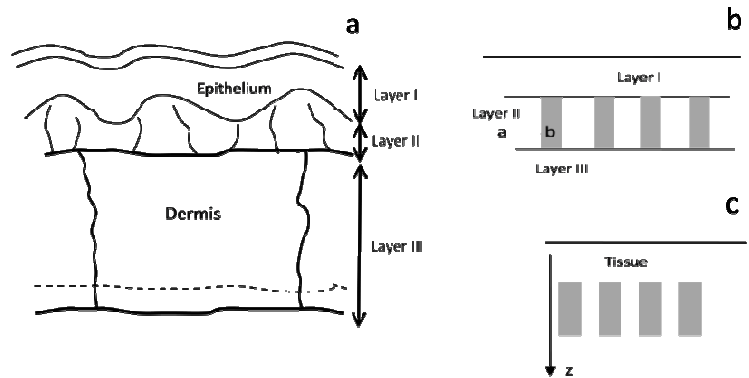

**Fig. 1** The logical transition from tissue microstructure (a) to heterogeneous dermis layer representation (b) to geometry allows evaluating the upper bound on the contrast ratio (c). Areas (a) and (b) of panel b are background tissue and capillary (defect), respectively. Reproduced from [9] with modifications.

Note that for consistency, we will use capital letters to indicate defect properties. E.g., $V$- defect volume, $Z$- defect depth, $D$ – defect height. However, we will use the traditional notation for the defect's optical properties (e.g., $\delta\mu_a$ is an incremental absorption coefficient associated with the defect). A point in the bulk of the tissue (3D) and on the surface of the tissue (2D) will be denoted (x,y,z) and <x,y>, respectively. We will also use "defect" and "inhomogeneity" interchangingly.

*2.3 Mathematical model*

The light propagation problem in homogeneous tissue can be solved exactly for certain geometries (e.g., slab, semi-space, or spheroid in diffuse approximation, slab and semi-space in Kubelka-Munk model (see, e.g. [11]). However, an arbitrary defect complicates things significantly, and we have to look for an approximate solution. A perturbation theory can be a helpful approach to find such an approximate solution: we start from the exact solution for the semi-space geometry and add the defect as a perturbation. Our perturbation approach consists of the following steps:

1. We solve the light propagation problem in homogeneous semi-infinite tissue (the radiant energy fluence rate $\varphi(\vec{\rho})$ (W/m$^2$)).

2. If we know the radiant energy fluence rate $\varphi(\vec{\rho})$ at some particular point $\vec{\rho}$, we can calculate additional (or incremental) absorbed optical power density $\delta\mu_a\varphi(\vec{\rho})$ (W/m$^3$) for some optical heterogeneity with the absorption coefficient $\mu_a+\delta\mu_a$ located at this point. If the volume of the heterogeneity is $V$, then the additional power absorbed at this heterogeneity will be $\delta\mu_a\varphi(\vec{\rho})V$ (W).

3. Alternatively, the heterogeneity can be considered a negative (or inverse) point source with power $-\delta\mu_a V \varphi(\vec{\rho})$ located at the point $\vec{\rho}$. The radiant energy fluence rate induced by such a source can be calculated exactly.

This problem can be analyzed using the diffuse approximation [12].

Step 1: For a semi-infinite medium with *wide beam diffuse illumination*, the total radiant energy fluence rate within the tissue far from the borders of the beam depends only on the depth $z$ (Eq. 6.88 in [12]):

$$\varphi_d(z) = \frac{4}{1-r_{10}} \frac{\exp(-\mu_{eff}z)}{1+h\mu_{eff}} \tag{1}$$

Where $\mu_{eff} = \sqrt{\mu_a/\delta}$, $\delta = 1/3\mu_{tr}$, $\mu_{tr} = \mu_a + \mu_s(1-g)$, $r_{10}$ - is the coefficient of reflection of diffuse light on the border of tissue and air ($r_{10}$ can be approximated using the relative index of refraction $n$: $r_{10}\approx 1-n^{-2}$), $h = 2\delta\frac{1+r_{10}}{1-r_{10}}$, $\mu_a$, $\mu_s$, and $g$ are coefficients of absorption, scattering, and anisotropy. Here without losing generality (we will be looking for the contrast ratio, which is dimensionless), we also assumed that the surface density of the incident light is 1 (W/m$^2$).

Similarly, we can solve the semi-infinite problem for *wide beam collimated illumination*. The difference here is the presence of collimated term, which dissipates proportionally to $\exp(-(\mu_a+\mu_s)z)$. For the biologically relevant case, $\mu_a \ll \mu_s$ we have an expression (Eq. 6.83 in [12]):

$$\varphi_c(z) = \frac{5-r_{10}}{1-r_{10}} \frac{\exp(-\mu_{eff}z)}{1+h\mu_{eff}} - 2\exp(-(\mu_a+\mu_s)z) \tag{2}$$

The advantage of the wide beam diffuse illumination scenario is that it allows obtaining closed-form expressions. Such as diffuse illumination is a quite realistic scenario (e.g., ambient light); we will limit ourselves to this scenario only. Further details for collimated illumination can be found in [9].

Step 2: The additional power absorbed at the inhomogeneity can be found by multiplication of Eq.1 on $V\delta\mu_a$,

Step 3: The diffuse source with power P in isotropic medium generates radiant energy fluence rate on the distance $\rho$ from the source

$$\varphi_s(\rho) = \frac{3P\mu_{tr}}{4\pi\rho} \exp(-\mu_{eff}\rho) \qquad (3)$$

Thus, we can represent our defect as the point source described by Eq.3, where power P is calculated on step 2 with a minus sign (negative source). To take into account the boundary conditions, we can use the diffusion dipole model [13,14] and in addition to the initial source located at depth Z, consider the second source (with opposite sign) located on the distance $2h+Z$ above the surface. In this case, total flux approximately satisfies realistic boundary condition for all r ($r = \sqrt{x^2 + y^2}$) and z=0 [15]

$$\varphi(r,z) - h\frac{\partial \varphi(r,z)}{\partial z} = 0 \qquad (4)$$

*2.4. Contrast ratio*

The contrast ratio provides a convenient way to analyze images. The contrast ratio at any point <x,y> on the surface of the tissue can be defined as

$$c(x, y) = \frac{\varphi_b - \varphi(x, y)}{\varphi_b} \qquad (5)$$

where $\varphi_b$ and $\varphi(x,y)$ are signals at some distant point (background) and any point of interest <x,y>, respectively. In our model, the flux on the surface of the tissue consists of two parts $\varphi(x,y)=\varphi+\varphi_s(x,y)$, where $\varphi$ is an unperturbed flux from a homogeneous tissue (does not depend on <x,y>) and $\varphi_s(x,y)$ is a flux caused by a negative point source (defect), respectively. We can take the unperturbed flux on the surface as a background ($\varphi_b=\varphi$). We cannot ignore the flux from the defect, $\varphi_s(x,y)$ near the inhomogeneity. Thus, we can write

$$c(x, y) = \frac{\varphi_b - \varphi(x, y)}{\varphi_b} = -\frac{\varphi_s(x, y)}{\varphi} \qquad (5^*)$$

3. **Results**

If the point source with power P is located at (0,0,Z), then the fluence rate at any point on the tissue surface (here we assume cylindrical coordinates) in the presence of mismatched boundary (Eq.4) will be:

$$\varphi_s(r) = \frac{3P\mu_{tr}}{4\pi}\left[\frac{\exp(-\mu_{eff}(Z^2+r^2)^{1/2})}{(Z^2+r^2)^{1/2}} - \frac{\exp(-\mu_{eff}((2h+Z)^2+r^2)^{1/2})}{((2h+Z)^2+r^2)^{1/2}}\right] \qquad (6)$$

Here again $\mu_{eff} = \sqrt{\mu_a/\delta}$, $\delta = 1/3\mu_{tr}$, $\mu_{tr} = \mu_a + \mu_s(1-g)$, $h = 2\delta\frac{1+r_{10}}{1-r_{10}}$, where, $r_{10}$ - is the coefficient of reflection of diffuse light on the border of tissue and air. As previously defined, r is the distance on the surface of the tissue from the projection of the defect to the surface ($r = \sqrt{x^2 + y^2}$).

As we already discussed, the power of the source $P$ can be assessed as $-\delta\mu_a \phi(Z) V$, where $\phi(Z)$ can be found from Eq.1 (diffuse illumination) or Eq.2 (collimated illumination). However, as mentioned before, we will limit our discussion to diffuse illumination cases only.

*3.1. Single-point defect*

We will keep considering the inhomogeneity located at *(0,0,Z)* and using cylindrical coordinates. Far from the inhomogeneity, its effect on flux on the surface is negligible. Thus, we can take the flux rate on the surface at this point as a background ($\phi_b=\phi_d(0)$). We cannot ignore the flux rate from the inhomogeneity, $\phi_s(r)$, near the defect. If we compare the background flux with the fluence rate on the surface in the presence of the inhomogeneity ($\phi(r)=\phi_d(0)+\phi_s(r)$), we can calculate the contrast ratio at any point on the surface of the tissue

$$c(r) = \frac{\varphi_b - \varphi(r)}{\varphi_b} = \frac{3\mu_{tr}\delta\mu_a V \exp(-\mu_{eff} Z)}{4\pi}\left[\frac{\exp(-\mu_{eff}(Z^2+r^2)^{1/2})}{(Z^2+r^2)^{1/2}} - \frac{\exp(-\mu_{eff}((2h+Z)^2+r^2)^{1/2})}{((2h+Z)^2+r^2)^{1/2}}\right] \quad (7)$$

Immediately above the heterogeneity *(r=0)* from Eq.7, we can get a compact expression

$$c(0) = \frac{3\mu_{tr}\delta\mu_a V \exp(-2\mu_{eff} Z)}{4\pi}\left[\frac{1}{Z} - \frac{\exp(-\mu_{eff} 2h)}{2h+Z}\right] \quad (8)$$

*3.2. Horizontally arranged defects*

We can notice that the contrast ratio is additive, which means that if we have several defects, we can add their effects. It follows from the definition of the contrast ratio (see Eq 5 and Eq.5*). Such as $c$ depends on $\varphi_s(x,y)$ in Eq.5* linearly, the contrast ratio is additive, which means that $c = \sum_i c_i$, where $c_i$ is the contrast ratio induced by the *i-th* defect.

We can use this observation to extend the point defect model to different defect geometries. In particular, let's assume that point defects with the length *dx* and cross-section area *S* are arranged in a line *(x,0, Z)*, where $-\infty<x<\infty$. Then, the contrast ratio on the surface of the tissue on the distance *y* from the projection of the "defect" line (lines *(x, y, 0)* and *(x,-y,0)* where $-\infty<x<\infty$) will be given by

$$c_{h,Z}(y) = \frac{3\mu_{tr}\delta\mu_a S \exp(-\mu_{eff} Z)}{4\pi} \times \int_{-\infty}^{\infty}\left[\frac{\exp(-\mu_{eff}(Z^2+y^2+x^2)^{1/2})}{(Z^2+y^2+x^2)^{1/2}} - \frac{\exp(-\mu_{eff}((2h+Z)^2+y^2+x^2)^{1/2})}{((2h+Z)^2+y^2+x^2)^{1/2}}\right]dx \quad (9)$$

Here for brevity and clarity, we introduced $c_{h,Z}(y) = c(x,y)|_{z=Z}$ for any *x*. We can simplify this expression if we notice the symmetry $-x \leftrightarrow x$

$$c_{h,Z}(y) = \frac{3\mu_{tr}\delta\mu_a S \exp(-\mu_{eff} Z)}{2\pi} \times \int_{0}^{\infty}\left[\frac{\exp(-\mu_{eff}(Z^2+y^2+x^2)^{1/2})}{(Z^2+y^2+x^2)^{1/2}} - \frac{\exp(-\mu_{eff}((2h+Z)^2+y^2+x^2)^{1/2})}{((2h+Z)^2+y^2+x^2)^{1/2}}\right]dx \quad (9*)$$

If we make substitutions $v = \mu_{eff} x$, $a = \mu_{eff}(Z^2 + y^2)^{1/2}$ and $b = \mu_{eff}((2h+Z)^2 + y^2)^{1/2}$ then we can rewrite Eq.9* as

$$c_{h,Z}(y) = \frac{3\mu_{tr}\delta\mu_a S \exp(-\mu_{eff} Z)}{2\pi} \int_0^\infty \left[ \frac{\exp(-(a^2+v^2)^{1/2})}{(a^2+v^2)^{1/2}} - \frac{\exp(-(b^2+v^2)^{1/2})}{(b^2+v^2)^{1/2}} \right] dv =$$
$$= \frac{3\mu_{tr}\delta\mu_a S \exp(-\mu_{eff} Z)}{2\pi} \int_0^\infty \left. \frac{\exp(-(t^2+v^2)^{1/2})}{(t^2+v^2)^{1/2}} \right|_b^a dv \qquad (9^{**})$$

If we change the order of operations under the integral, we can notice that using substitution $v = t\,sh(u)$, the integral can be transformed into a canonical form

$$\int_0^\infty \frac{\exp(-(t^2+v^2)^{1/2})}{(t^2+v^2)^{1/2}} dv \xrightarrow{v=t\times sh(u)} \int_0^\infty \exp(-t\times ch(u))\,du = K_0(t) \qquad (10)$$

Here $K_0(t)$ is the modified Bessel function of the second kind. Thus, the explicit expression for the contrast ratio of the linear defect buried on the depth $Z$ measured on the distance $y$ from it horizontally can be written as

$$c_{h,Z}(y) = \frac{3\mu_{tr}\delta\mu_a S \exp(-\mu_{eff} Z)}{2\pi} K_0(t)\Big|_b^a =$$
$$= \frac{3\mu_{tr}\delta\mu_a S \exp(-\mu_{eff} Z)}{2\pi} (K_0(\mu_{eff}(Z^2+y^2)^{1/2}) - K_0(\mu_{eff}((2h+Z)^2+y^2)^{1/2})) \qquad (11)$$

For example, the contrast ratio at any point above the defect will be given by $c_{h,Z}(0)$.

*3.3. Vertically arranged defects*

A similar approach can be developed for the vertically- arranged defects. In this case, the vessel can be approximated as a segment with a length $D$ (defect height): $z=[Z,Z+D]$. Similarly to Eq.9, we can write taking into account rotational symmetry in the $xy$ plane:

$$c_{v,Z,l}(r) = \frac{3\mu_{tr}\delta\mu_a S}{4\pi} \int_Z^{Z+D} \exp(-\mu_{eff} z) \left[ \frac{\exp(-\mu_{eff}(z^2+r^2)^{1/2})}{(z^2+r^2)^{1/2}} - \frac{\exp(-\mu_{eff}((2h+z)^2+r^2)^{1/2})}{((2h+z)^2+r^2)^{1/2}} \right] dz \qquad (12)$$

This integral cannot be evaluated analytically. However, we can evaluate it numerically in realistic cases.

We have numerically evaluated single point, horizontally- and vertically-arranged defects using Eq.7, 11, and 12, respectively. For these purposes, we split the single point defect volume $V$ as $V=dXdYdZ$ and assumed that the $dX=dY=dZ=20\mu m$. In the case of horizontally- and vertically- arranged defects, we integrated over $x$ and $z$, respectively, keeping the same cross-section area ($S=20\mu m \times 20\mu m$).

If we take realistic assumptions for mucosa: $\mu_a$=0.001mm$^{-1}$, $\mu_s$'=5mm$^{-1}$ (at 532nm), $\delta\mu_a$= 28mm$^{-1}$ (the whole blood with 70% oxygenation at 532nm), and $n$=1.33 we can calculate the contrast ratio for various defect geometries.

While for the single point and the horizontally-arranged defects, the outcome depends on the volume (or cross-section) and the defect depth, $Z$ only, the vertically arranged defect outcome depends on the defect height, $D$. Thus, to compare all three scenarios, firstly, we need to understand that dependence.

In Fig 2, one can see the dependence of the maximal contrast ratio (just above the defect, $r=0$) in the case of the vertical defect as a function of the defect height, $D$. We evaluated it for the entire thickness of the dermis layer (2mm, panel A) and the reasonable estimation of the nurturing capillary height (0.2mm, panel B).

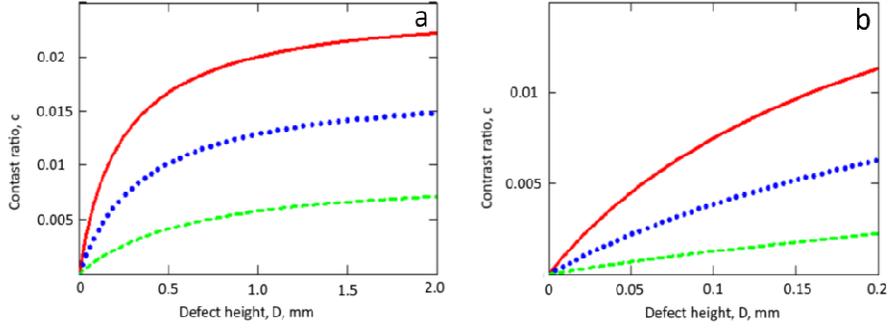

**Fig. 2** Vertically-arranged defects. The dependence of the contrast ratio above the defect ($r=0$) as a function of the defect height, $D$. for $Z=0.1$mm (red solid line), 0.2mm (blue dotted line), and 0.5mm (green dashed line). Panel a: full physiological range (dermis, 2mm). Panel b: zoomed-in area (papillary dermis, 0.2mm)

In Fig 3, one can see a comparison of the dependence of the contrast ratio on the distance for the point defect (left pane), horizontally- and vertically arranged defects (central and right pane, respectively). Defect depth was $Z=0.1$mm (solid red line), 0.2mm (blue dotted line), and 0.5mm (green dashed line). In the case of a vertical defect, the defect height was set at $D=0.2$mm (nurturing capillaries).

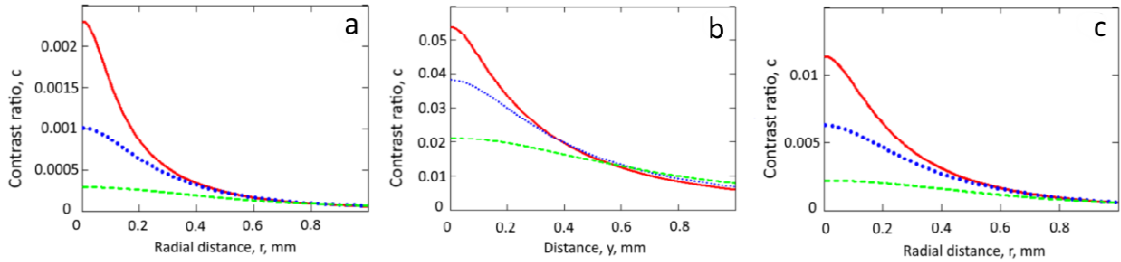

**Fig. 3** The radial dependence of the contrast ratio for the point defect (panel a), horizontally- and vertically arranged defects (panels b and c, respectively). Defect depth, $Z=0.1$mm (red solid line), 0.2mm (blue dotted line), and 0.5mm (green dashed line). In the case of a vertical defect, the defect height, $D=0.2$mm

## 4.     Discussion

We developed the single point subsurface defect model in turbid media with mismatched boundaries and extended it to horizontally- and vertically-arranged linear inhomogeneities. In particular, we have obtained an explicit analytical expression for the infinite linear defect buried at a certain depth, which allows direct experimental verification.

The results are in agreement with previously published numerical [1] and Monte Carlo [7] calculations. The results are also in agreement with the fact that nail fold (horizontally-arranged) capillaries are clearly visible during capillaroscopy.

The expression for the single point defect (Eq.7) can be considered an analog of the point spread function (PSF) for the reflectance imaging geometry.

One of the advantages of the metric we used (the contrast ratio) is its additivity. This property allows aggregating individual impacts for groups of defects. Thus, for example, we can further explore the contrast ratio's addictiveness and integrate Eq.11 over the cross-section of the blood vessel lumen. However, the utility and validity of such an approach require further justification as the developed approach is the first-order perturbation technique that does not consider the interaction between defects, which may become important.

Nevertheless, some geometric considerations can be used to take into account the interaction between defects. The interaction from nearby defects can be accounted for under the assumption of isotropic light distribution in the tissue. In this case, we can assume that the light is coming either from obstructed areas (nearby defects) or unobstructed areas (unperturbed tissue) and calculate the solid angle of the obstructed areas assuming that the light is predominantly absorbed there. While it is difficult to assess the obstructed areas' solid angle accurately, we can do it roughly in a simple lattice model. Let's consider a cubic lattice. In this case, each cell has six nearest neighbors. If we consider the line of the absorbing objects in such geometry, then each defect has two high-absorbing neighbors and four low-absorbing neighbors. Thus, we can expect that the contrast ratio can be corrected by the factor of 4/6=2/3. Even more accurate calculations can be based on the relative light attenuation of a high-absorbing neighbor vs. a low-absorbing neighbor. For example, the high-absorbing neighbors can be described by an additional factor $exp(-\delta\mu_a l)$. Here, $l$ is the size of the lattice, e.g., a typical defect size or the vessel diameter. Let's assume that light intensity from the unobstructed area is $i_o$ and the light intensity in the defect is $i$. Then, we can write the equation for any defect on the line as

$$i = \frac{4}{6}i_0 + \frac{2\exp(-\delta\mu_a l)}{6}i \tag{13}$$

Solving this equation for $i$, we can find the overall 1D correction factor $\gamma_{1D}$ ($i = \gamma_{1D} i_0$)

$$\gamma_{1D} = \frac{2/3}{1 - 1/3\exp(-\delta\mu_a l)} \tag{14}$$

Similarly, for a 2D absorption defect (e.g., a pool of blood), we can expect that the light will be obstructed from 4 out of 6 directions. Thus, the overall 2D correction factor can be estimated as

$$\gamma_{2D} = \frac{1/3}{1 - 2/3\exp(-\delta\mu_a l)} \tag{15}$$

The effect of the defect height is two-fold (see left vs. right pane of Fig 3). Firstly, with the increase of height from 0.02mm to 0.2mm, the contrast increased 5-fold. Secondly, the "width of the influence," which can be measured as a half-width at half maximum (HWHM), also increases with the defect height.

The developed approach is based on the diffuse approximation. Thus, it carries all its limitations. In particular, the diffusion scattering approach is questionable on the distances shorter than inversed reduced scattering coefficient $1/\mu'_s$ from the boundary [12]. Thus, in our illustration example ($\mu_s'$=5mm$^{-1}$ (at 532nm)), if the defect is closer than

200μm, our approach's validity is questionable. However, in the realistic skin at 633nm ($\mu_s'$=14mm$^{-1}$ and 9mm$^{-1}$ for stratum corneum and living epidermis, respectively [10]), this distance is 100μm and less, which is smaller than the normal epidermis thickness. Thus, for the normal skin, our approximation should be valid. However, rigorous comparison with the gold standard (Monte Carlo calculations) and identification of the area of applicability are necessary. In future work, we also plan to verify theoretical predictions in phantom experiments.

## 5. Conclusions

We developed the single point subsurface defect model in turbid media with mismatched boundaries and extended it to horizontally- and vertically-arranged linear inhomogeneities. The developed approach can be used for quick rough estimates while designing and optimizing imaging systems.